# The incompleteness of the experimental searches for vector-like leptons at CERN

Feyza Baspehlivan[1], Burak Dagli[1], Hatice Duran Yildiz[2], and Saleh Sultansoy[1,3]

[1]TOBB University of Economics and Technology, Ankara, Türkiye
[2]Ankara University, Institute of Accelerator Technologies, Ankara, Türkiye
[3]ANAS Institute of Physics, Baku, Azerbaijan

**Abstract**

There are strong phenomenological arguments favoring the existence of vector-like leptons and quarks in nature. In spite of extensive studies conducted in search of vector-like quarks, there are only a limited number of experimental studies on vector-like leptons. Moreover, these searches do not include all possible decay modes of vector-like leptons. Therefore, the analyses done so far are incomplete. In this letter, we highlight decay channels that are not covered by different experimental analyses, with a focus on L3, ATLAS, and CMS results. We argue that experimental analyses should be redone considering these shortcomings.

*Introduction.* — Despite the completion of the electroweak part of the Standard Model (SM) of elementary particles by the discovery of the Higgs Boson [1, 2], there remain a lot of unresolved issues that require an explanation, such as the mass and mixing patterns of the SM fermions (according to Steve Weinberg, the observed pattern of quark and lepton masses is one of the most important mysteries to be solved [3]), left-right asymmetry, huge difference between electroweak and Planck scales (for a more complete list, see i.e., reviews [4, 5]). For this reason, numerous new models beyond the SM (BSM) have been proposed, and these models predict a large number of new particles. Vector-like leptons (VLL) and quarks (VLQ) are one of the most intriguing phenomena that have the potential to explain SM fermions' masses and mixing patterns according to the Flavor Democracy (FD) Hypothesis (see [6] and references therein). Vector-like fermions are suggested by a number of popular BSM models: $E_6$ Grand Unification Theory (GUT), composite models, and See-Saw III are some examples. Besides, VLLs could provide solutions for several anomalies in particle physics [7] (see for example [8] for the Cabibbo Angle Anomaly and [9, 10] for the muon g-2 anomaly).

Vector-like fermions exhibit the property that their left- and right-handed components transform in the same way under the symmetry group of theory. It is worth reminding that in the SM, left-handed fermions form SU(2) doublets, whereas the right-handed ones are singlets. Therefore, the mass terms of vector-like fermions are not prohibited by gauge symmetry, unlike the fourth SM family fermions that have been excluded based on recent findings related to the Higgs boson.

Iso-doublet VLL partners of all three SM families are predicted by the $E_6$ GUT [11, 12]. The first family fermion sector of the $E_6$-induced model has the following $SU_c(3) \times SU_w(2) \times U_Y(1)$ structure:

$$\begin{pmatrix} u_L \\ d_L \end{pmatrix} u_R \; d_R \; D_L \; D_R \begin{pmatrix} \nu_{e_L} \\ e_L \end{pmatrix} \nu_{e_R} \; e_R \begin{pmatrix} N_{e_L} \\ E_L \end{pmatrix} \begin{pmatrix} N_{e_L} \\ E_R \end{pmatrix} \mathcal{N}_e$$

where D denotes a new quark with Q = -1/3, $N_e$ and E are new neutral and charged leptons, and $\mathcal{N}_e$ is heavy Majorana neutral lepton.

Flavor Democracy is another motivation for VLLs. The FD Hypothesis predicts the existence of iso-singlet or iso-doublet vector-like leptons as well as iso-singlet vector-like quarks with Q = -1/3 and provides a hierarchical structure among elementary particles [13-15]. Initially, this hypothesis was proposed for the SM with three families. However, the observation that the mass of the top quark is much higher than the mass of the b quark disproved this hypothesis. Later, it was shown that the FD hypothesis can be survived by adding a fourth family to the SM [16-18]. Unfortunately, the measurements concerning the properties of the Higgs Boson exclude this possibility. Let us emphasize that the FD hypothesis, which is quite natural in the SM framework, can be survived by introducing vector-like leptons and quarks [13-15]. In other words, FD calls for vector-like partners of the SM leptons and quarks.

From a particle phenomenology point of view, vector-like quarks and vector-like leptons have the same status. However, although extensive studies have been conducted on the search for VLQs, there are a limited number of experimental studies on VLLs. Unfortunately, these searches do not include all possible decay modes of VLLs (see [15] for details). Therefore, the analyses done so far are incomplete.

All possible decay modes of iso-singlet and iso-doublet VLLs are considered in [15] (below this option is named General Model – GM). The model [19, 20] used in recent ATLAS and CMS experimental analyses corresponds to a specific case where some mixings were set to zero and the equivalence of charged and neutral VLL masses has been assumed (below this option is named Restricted Model – RM). In this letter, the experimental results of L3, ATLAS, and CMS collaborations on the search for charged and neutral VLLs are discussed, and the missing decay channels for each experiment are laid out.

The decay behavior of vector-like leptons is determined by the relationship between the mixing angles and the masses of charged and neutral vector-like leptons. It is natural to assume that each SM family has its own vector-like partners; therefore, one deals with 6×6 mass matrices both for charged and neutral (Dirac) leptons. However, if neutrinos and vector-like neutral leptons have Majorana masses, the corresponding mass matrix has 12×12 dimensions. Since vector-like leptons' masses and mixings are not known, we are faced with a very complex situation. For this reason, we consider the case where neutral leptons have only Dirac masses and intra-family mixings of vector-like leptons with SM leptons are dominant.

In the next two sections, we present formulas for the decays of the first family-related vector-like leptons, the decay processes for second and third family-related VLLs are the same except that they are replaced by particles belonging to the corresponding family. Let us mention that the ATLAS and CMS experiments mainly focus on third family-related VLLs and their analyses are based on the RM, which does not include a number of decay modes, while it is quite possible that these decay modes are dominant ones. The model [15], which is the basis of our consideration, is more general and is reduced to the RM for definite parameter sets.

The decay width formulas given below are taken from [15] (details of the model and notations can be found in this reference). The following notations are used: $M_E$ and $M_{Ne}$ are masses of first SM family-related charged and neutral vector-like leptons, respectively; $a_E = g^2 M_E^2/2m_W^2$, $a_N = g^2 M_N^2/2m_W^2$; $r_X^E = M_X^2/M_E^2$ and $r_X^N = m_X^2/M_N^2$ ($X = W, Z, H$); $g = \sqrt{4\pi\alpha}/sin\theta_W$, where $\alpha$ is the fine-structure constant and $\theta_W$ is the Weinberg angle. Finally, s and c stand for the sinus and cosine of mixing angles arising as a result of the transition from the SM basis, denoted as superscript zero, to the mass basis (for details, see [15]):

$$e_L^0 = c_L^E e_L + s_L^E E_L, \qquad e_R^0 = c_R^E e_R + s_R^E E_R$$
$$E_L^0 = -s_L^E e_L + c_L^E E_L, \qquad E_R^0 = -s_R^E e_R + c_R^E E_R$$
$$\nu_L^0 = c_L^N \nu_L + s_L^N N_L, \qquad \nu_R^0 = c_R^N \nu_R + s_R^N N_R$$
$$N_L^0 = -s_L^N \nu_L + c_L^N N_L, \qquad N_R^0 = -s_R^N \nu_R + c_R^N N_R$$

*Decays of Iso-singlet Vector-like Leptons.*—It should be noted that neutral iso-singlet leptons are not included in the RM (this corresponds to $s_L^N = s_R^N = 0$ case in the GM). Most likely, the authors of [19, 20] assumed that neutrinos have no right-handed components (hence, iso-singlet neutral vector-like leptons do not exist). Let us emphasize that right-handed neutrinos should be included in the SM since, according to quark-lepton symmetry, there are counterparts of right-handed components of up quarks, and observation of neutrino oscillations confirms this statement.

The decays of first family-related VLLs are caused by E-e and $N_e$-$\nu_e$ mixings, and the corresponding decay width equations for the iso-singlet charged vector-like lepton are as follows

$$\Gamma(E^- \rightarrow W^- \nu) = \frac{M_E}{32\pi} a_E (c_L^N)^2 (s_L^E)^2 (1 - r_W^E)^2 (1 + 2r_W^E) \tag{1}$$

$$\Gamma(E^- \rightarrow Z e^-) = \frac{M_E}{64\pi} a_E (c_L^E)^2 (s_L^E)^2 (1 - r_Z^E)^2 (1 + 2r_Z^E) \tag{2}$$

$$\Gamma(E^- \rightarrow H e^-) = \frac{M_E}{64\pi} a_E (c_L^E)^2 (s_L^E)^2 (1 - r_H^E)^2 \tag{3}$$

If E is much heavier than W, Z and H bosons branching ratios become BR$(E^- \rightarrow W^- \nu) = 0.5$ and BR$(E^- \rightarrow Z e^-) =$ BR$(E^- \rightarrow H e^-) = 0.25$.

Equations (1) - (3) coincide with the corresponding equations from [19, 20] if $s_L^N = 0$ (since neutral VLL is not included in the iso-singlet RM), $c_L^E \cong 1$ and $\epsilon = \frac{g}{\sqrt{2}} \frac{M_E}{M_W} s_L^E$, where $\epsilon$ is mixing Yukawa coupling (see [19] for details).

Decay width formulas for iso-singlet neutral vector-like leptons are

$$\Gamma(N \rightarrow W^+ e^-) = \frac{M_N}{32\pi} a_N (c_L^E)^2 (s_L^N)^2 (1 - r_W^N)^2 (1 + 2r_W^N) \tag{4}$$

$$\Gamma(N \rightarrow Z \nu) = \frac{M_N}{64\pi} a_N (c_L^N)^2 (s_L^N)^2 (1 - r_Z^N)^2 (1 + 2r_Z^N) \tag{5}$$

$$\Gamma(N \rightarrow H \nu) = \frac{M_N}{64\pi} a_N (c_L^N)^2 (s_L^N)^2 (1 - r_H^N)^2 \tag{6}$$

Similar to the case of the E, if the mass of the neutral vector-like lepton N is much larger than the masses of the W, Z, and H bosons, the branching ratios become BR$(N \rightarrow W^+ e^-) = 0.5$ and BR$(N \rightarrow Z \nu) =$ BR$(N \rightarrow H \nu) = 0.25$. Let us remind that the RM does not include neutral iso-singlet leptons.

*Decays of Iso-doublet Vector-like Leptons.*—As mentioned in the introduction, iso-doublet vector-like leptons are predicted by the Flavor Democracy Hypothesis and the $E_6$ model. Similar to the iso-singlet case, the RM does not include a number of possible decay modes in the iso-doublet case either because of the lack of right-handed components of neutrinos. The second important point is degeneracy of charged and neutral VLLs in the RM, which is the result of assuming that only one iso-doublet exists in this model. Let us emphasize that $E_6$ GUT predicts the existence of three iso-doublets, one doublet per SM family, and in this case, there is no reason for degeneracy. As a result, a number of new decay channels should be taken into account as well.

If E is the lightest VLL, E will decay only through mixings with the first SM family leptons and corresponding decay width formulas are

$$\Gamma(E^- \rightarrow W^- \nu) = \frac{M_E}{32\pi} a_E [(c_L^N s_L^E - c_L^E s_L^N)^2 + (c_R^E)^2 (s_R^N)^2](1 - r_W^E)^2 (1 + 2r_W^E) \tag{7}$$

$$\Gamma(E^- \rightarrow Z e^-) = \frac{M_E}{64\pi} a_E (c_R^E)^2 (s_R^E)^2 (1 - r_Z^E)^2 (1 + 2r_Z^E) \tag{8}$$

$$\Gamma(E^- \rightarrow H e^-) = \frac{M_E}{64\pi} a_E (c_R^E)^2 (s_R^E)^2 (1 - r_H^E)^2 \tag{9}$$

In the $M_E \gg m_{W,Z,H}$ case, branching ratios become BR$(E^- \to W^- \nu)$ = 0.5 and BR$(E^- \to Z e^-)$ = BR$(E^- \to H e^-)$ = 0.25 if $s_L^E = s_L^N$ and $s_R^E = s_R^N$.

It should be noted that the $E^- \to W^- \nu$ decay width is zero in the RM, which corresponds to $s_L^E = s_L^N$ and $s_R^N = 0$ in Equation (7). The reason the $s_R^N$ is zero is the absence of right-handed neutrinos in this model (as highlighted above, right-handed neutrinos should be included in the game). The reason $s_L^E$ is equal to $s_L^N$ is that the masses of E and N are equal in this model (as mentioned above, this applies when there is only one vector-like doublet). Decays into the Z and H channels overlap with equations 2.26 and 2.27 in [19] if $\epsilon = \frac{g}{\sqrt{2}} \frac{M_E}{M_W} s_R^E$.

If N is the lightest VLL, its decay width formulas are given by

$$\Gamma(N \to W^+ e^-) = \frac{M_N}{32\pi} a_N \left[ (c_L^N s_L^E - c_L^E s_L^N)^2 + (c_R^N)^2 (s_R^E)^2 \right] (1 - r_W^N)^2 (1 + 2r_W^N) \tag{10}$$

$$\Gamma(N \to Z \nu) = \frac{M_N}{64\pi} a_N (c_R^N)^2 (s_R^N)^2 (1 - r_Z^N)^2 (1 + 2r_Z^N) \tag{11}$$

$$\Gamma(N \to H \nu) = \frac{M_N}{64\pi} a_N (c_R^N)^2 (s_R^N)^2 (1 - r_H^N)^2 \tag{12}$$

In the $M_N \gg m_{W,Z,H}$ case, branching ratios become BR$(N \to W^+ e)$ = 0.5 and BR$(N \to Z\nu)$ = BR$(N \to H\nu)$ = 0.25 if $s_L^E = s_L^N$ and $s_R^E = s_R^N$. The variation of branching ratios with respect to mass corresponding to this case is shown in Figure 2. If $s_L^E = s_L^N$ and $s_R^E = 0$, only neutral current decay modes survive. However, if the $s_R^N = 0$, only $N \to W^+ e$ channel remains.

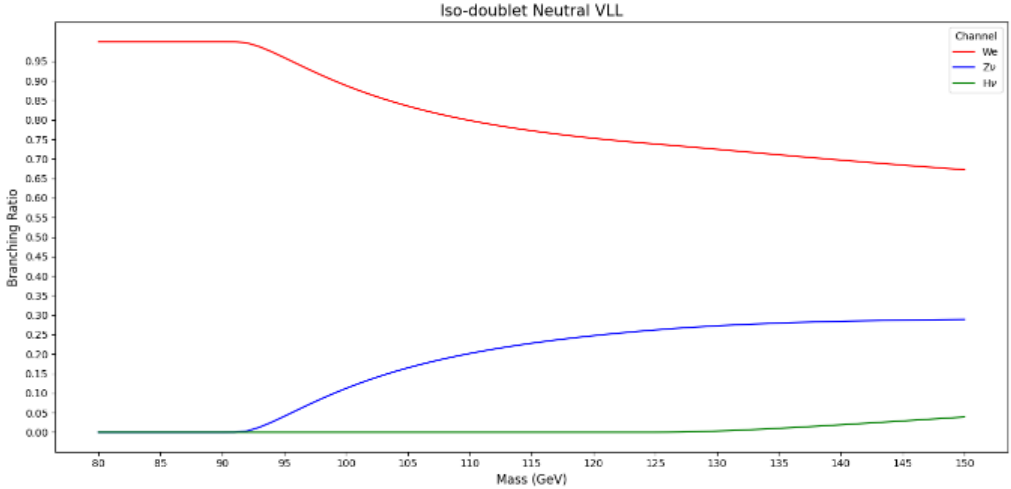

**Figure 1:** Branching Ratios of iso-doublet neutral VLL

Our formulas for neutral vector-like lepton decays overlap with 2.28 and 2.29 in [19] if $s_R^N = 0$, which result in zero decay width for neutral current decays (let us remind that the authors of [19] assumed that there are no right-handed components of neutrinos). Concerning the surviving charged decay mode Equation (10) coincides with equation 2.28 in [19] if $\epsilon = \frac{g}{\sqrt{2}} \frac{M_E}{M_W} \sqrt{(c_L^N s_L^E - c_L^E s_L^N)^2 + (c_R^N)^2 (s_R^E)^2}$.

If the E is heavier than $M_N + m_W$, a new decay channel $E^- \to W^- N$ becomes dominant, and the corresponding decay width is as follows

$$\Gamma(E^- \to W^- N) = \frac{M_E}{32\pi} a_E \left[ (1 - r_N^E - r_W^E)^2 - 4 r_N^E r_W^E \right]^{1/2} \left[ (1 - r_N^E)^2 + (1 + r_N^E - 6\sqrt{r_N^E}) r_W^E - 2(r_W^E)^2 \right] \tag{13}$$

In this formula, terms that are proportional to the square of mixing angles are neglected. The reason for the dominance of this channel is that the decays of E to $W\nu_e$, Ze, and He are suppressed by the square of the mixing angles.

On the other hand, if the mass of $E^-$ is between the mass of N and the sum of the masses of N and the W boson, W boson can be created virtually, and then $E^-$ will decay through the channels $E^- \to N\, l^- \nu_l$ and $E^- \to N\, q\bar{q}$. The dominance of $E \to N W^*$ decay mode depends on relation between values of mixing angles in Equations (7) - (9) and E - N mass difference. For illustration, dependence of decay widths on $M_N$ is shown in Figure 2 for E with 110 GeV mass. For instance, if $s_R^E$ and $s_R^N$ are less than 0.01 and $M_E - M_N$ is above 40 GeV, the $NW^*$ channel becomes dominant.

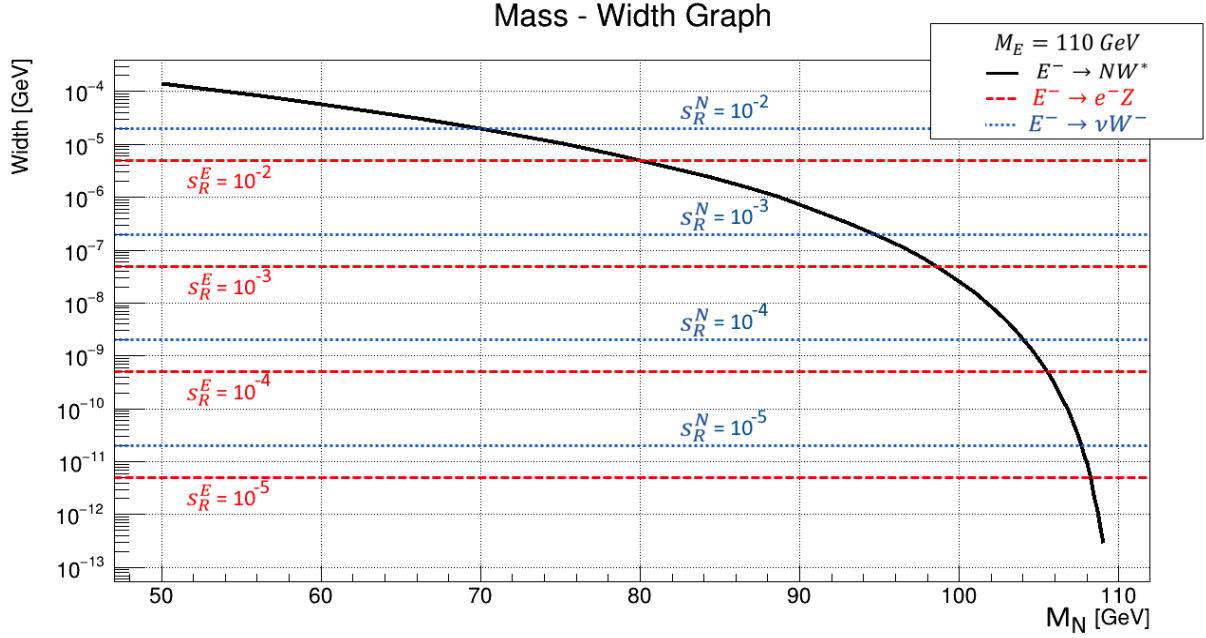

**Figure 2:** Illustration of charged VLL decay channel dominance for $M_N < M_E < M_E + m_W$ case.

It should be emphasized that N produced as a result of the decay of E will consequently decay into the channels given in Equations (10) - (12).

If N is heavier than E, a new decay channel of N emerges. If N has a mass greater than the sum of the masses of W and E, $N \to W^+ E^-$ decay channel will be dominant. The corresponding decay width formula is

$$\Gamma(N \to W^+ E^-) = \frac{M_N}{32\pi} a_N \left[(1 - r_E^N - r_W^N)^2 - 4 r_E^N r_W^N\right]^{1/2} \left[(1 - r_E^N)^2 + (1 + r_E^N - 6\sqrt{r_E^N})\, r_W^N - 2(r_W^N)^2\right] \quad (14)$$

In this formula, terms that are proportional to the square of mixing angles are neglected. The reason for the dominance of this channel is that the decays of N to We, $Z\nu_e$, and $H\nu_e$ are suppressed by the square of the mixing angles.

If the mass of N is lighter than the sum of W and E masses, N may decay through virtual W boson, namely, $N \to E^- l^+ \nu_l$ and $N \to E^- q\bar{q}$ channels. The dominance of $N \to E W^*$ decay mode depends on relation between values of mixing angles in Equations (10) - (12) and the N - E mass difference. It should be highlighted that E produced as a result of the decay of N will consequently decay into the channels given in Equations (7) - (9).

Below, we discuss the experimental results of searches for vector-like leptons. The strongest limits at lepton colliders were obtained by LEP L3 collaboration [21]. Recent searches at the LHC have been performed mostly on third family-related VLLs [22, 23]. First and second family-related VLL studies were done years ago by ATLAS collaboration with $\sqrt{s}$ = 8 TeV [24].

*LEP L3 Data.*—At LEP pair production of charged and neutral vector-like leptons, predicted by $E_6$ GUT, was investigated using L3 data [21]. The lower exclusion mass limits for neutral and charged vector-like leptons obtained in the analysis are shown in Tables 1 and 2, respectively.

**Table 1:** 95% CL lower mass limits for pair produced neutral vector-like leptons

| Decay mode | Lower mass limit (GeV) |
|---|---|
| We | 102.6 |
| Wμ | 102.7 |
| Wτ | 99.3 |

The decay widths of neutral vector-like leptons for the iso-doublet case are given in Equations (10) - (12). As seen from Table 1, neutral VLL decays into light leptons only via the charged current has been studied at L3, since they assumed that more than 90% of the VLLs decay through the W channel if their masses lie below 100 GeV (see also Figure 1). However, this assumption applies if $s_R^N = s_R^E$ and $s_L^N = s_L^E$. If $s_R^N$ is greater than $s_R^E$ and the $(c_L^N s_L^E - c_L^E s_L^N)$ difference is close to zero, according to Equations (10) - (12), the neutral current decay mode N → Zν becomes dominant. In this case, the limits presented in Table 1 are not valid and *the neutral current decay mode should be studied separately*. In a detector, the signal process involves *two Z bosons and large missing transverse momentum*. Let us mention that channel N → Hν is kinematically forbidden for N's mass values that can be scanned at LEP energies.

**Table 2:** 95% CL lower mass limits for pair produced charged vector leptons

| Decay mode | Lower mass limit (GeV) |
|---|---|
| νW | 101.2 |
| NW* | 102.1 |

The decay widths of charged vector-like leptons for the iso-doublet case are given in Equations (7) - (9). For unstable charged VLLs, only charged current decay modes have been examined and separate analyses have been conducted for the decay to neutrino and neutral VLL. However, depending on the mixing angles, neutral current decay modes may be dominant in the first scenario (Wν). Moreover, in the case of $s_L^E = s_L^N$ and $s_R^N = 0$, only neutral current decay modes are survived. Therefore, the limit presented in Table 2 is not inclusive, and *the neutral current decay mode should be taken into account*. In a detector, this process would be observed as *two Z bosons and a pair of oppositely charged leptons*. Let us mention that channel E → Hl is kinematically forbidden for E's mass values that can be scanned at LEP energies.

The second scenario investigated for charged VLL is the decay to stable neutral VLL (see Figure 5 in [21]). The corresponding decay width is given by Equation (13). Let us note that both neutrino and neutral VLL decay channels take place if $M_E > M_N$. Competition between two decay modes depends on mass differences and mixing angles as shown in the example given in Figure 2.

*ATLAS: 8 TeV Data.*—At the LHC, the production of the first and second family-related VLLs was studied at $\sqrt{s}$ = 8 TeV by the ATLAS Collaboration [24]. Only pair production of charged VLLs with subsequent decay of one of the vector-like lepton to a Z boson and either an electron or muon have been considered. The corresponding Feynman diagram is shown in Figure 3.

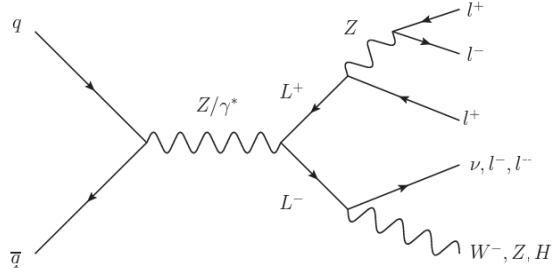

**Figure 3:** Feynman diagram for process analyzed by ATLAS [24]

The obtained exclusion limits are presented in Table 3. The second row corresponds to electron-only mixing and the third row corresponds to muon-only mixing cases.

**Table 3:** VLL mass exclusion limits

| Family | Channel | Exclusion limits (GeV) |
|---|---|---|
| First | E → Ze | 129-144 & 163-176 |
| Second | M → Zμ | 114-153 & 160-168 |

Decay widths of charged VLLs to Wν, Zl and Hl channels are given in Equations (7) - (9). One can see that if $s_R^E = 0$, neutral current decay modes vanish and only charged current decay modes survive. Therefore, the exclusion limits given in Table 3 are not applicable for the $s_R^E = 0$ case and *analyses should be redone keeping in mind the missing charged current decay channels*, namely, E → W $\nu_e$ and M → W $\nu_\mu$.

If charged VLLs are heavier than neutral partners, additional decay modes E → W $N_e$ and M → W $N_\mu$ emerge for the iso-doublet case (see Equation (13)). Moreover, produced neutral leptons are subject to further decay according to Equations (10) - (12). These chains should also be taken into consideration for a complete analysis.

*ATLAS and CMS: 13 TeV Data.*—Recent VLL search analyses of ATLAS and CMS Collaborations are devoted to the third SM family-related vector-like leptons (see References [23] and [22], respectively). This is a consequence of the assumption that VLLs are dominantly mixed with the third SM family leptons. This assumption, which originates from the notion that the hierarchical structure of the CKM matrix also holds for mixings of vector-like leptons, is incorrect. The structure of the PMNS matrix also confirms this statement (for details see [15]).

Concerning production processes in doublet case both collaborations analyzed pp → $\tau'^+\tau'^-/\nu'\bar{\nu}'/\tau'\nu'$ with subsequent decay modes $\tau' \to Z\tau$ or $H\tau$ and $\nu' \to W\tau$. In the singlet case, only pair production of charged lepton $\tau'$ is considered with subsequent decays $\tau' \to Z\tau$, $H\tau$ or $W\nu$ by CMS. Let us mention that the ATLAS study includes only iso-doublet case. Both collaborations assume degenerate masses of $\tau'$ and $\nu'$. As a result, CMS excluded masses below 1045 GeV for the iso-doublet case and between 125 and 150 GeV for the iso-singlet case. ATLAS excluded masses between 130 and 900 GeV for the iso-doublet case (as seen in Figure 4, lower limit in the ATLAS results has a technical character: the lowest mass point in the analysis has been chosen as 130 GeV instead of 100 GeV). Let us remind that L3 excluded masses below 100 GeV.

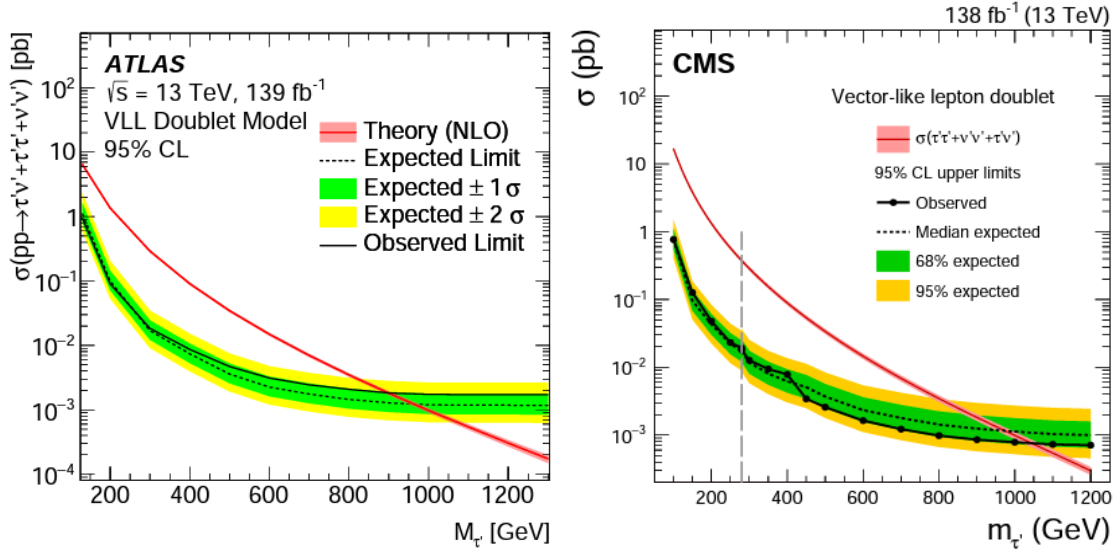

**Figure 4:** The 95% CL exclusion limits on the production cross-section as a function of VLL mass (left figure corresponds to ATLAS [23] and right to CMS [24] results).

For the iso-doublet VLLs decay width formulas (7) - (14) are applicable with the replacement of E, N, e and ν by τ′, ν′, τ and ν, respectively. With similar replacements in formulas (1) - (3), one can obtain iso-singlet τ′ decay widths. As mentioned above, for iso-doublet leptons ATLAS and CMS consider τ′ → Zτ or Hτ and ν′ → Wτ decay modes. Concerning the charged VLL, τ′ → Wν channel, which is not covered by their analysis, disappears only if $s_L^E = s_L^N$ and $s_R^N = 0$ (see Equation (7)). However, different from the RM, in general case there is no reason for these restrictions. Moreover, if $s_R^E = 0$ neutral channel modes disappear and one deals with τ′ → Wν decay mode only.

Concerning the neutral VLL, ν′ → Zν and ν′ → Hν channels, which are not considered by the ATLAS and CMS analyses, disappear only if $s_R^N = 0$ (see Equations (11) - (12)). Again, different from the RM, in general case there is no reason for this restriction. Furthermore, if $s_L^E = s_L^N$ and $s_R^E = 0$, charged channel mode disappears, and one should analyze neutral channel modes separately.

Cross-sections for pair and associate productions of iso-doublet vector-like leptons at the LHC are shown in Figure 5 [15], where cross-sections for associate E and N productions are calculated assuming $M_E = M_N$. It is seen that the $E^+N$ cross-section exceeds the other three approximately by a factor of 4. Certainly ATLAS and CMS exclusion limits are not applicable to non-degenerate case.

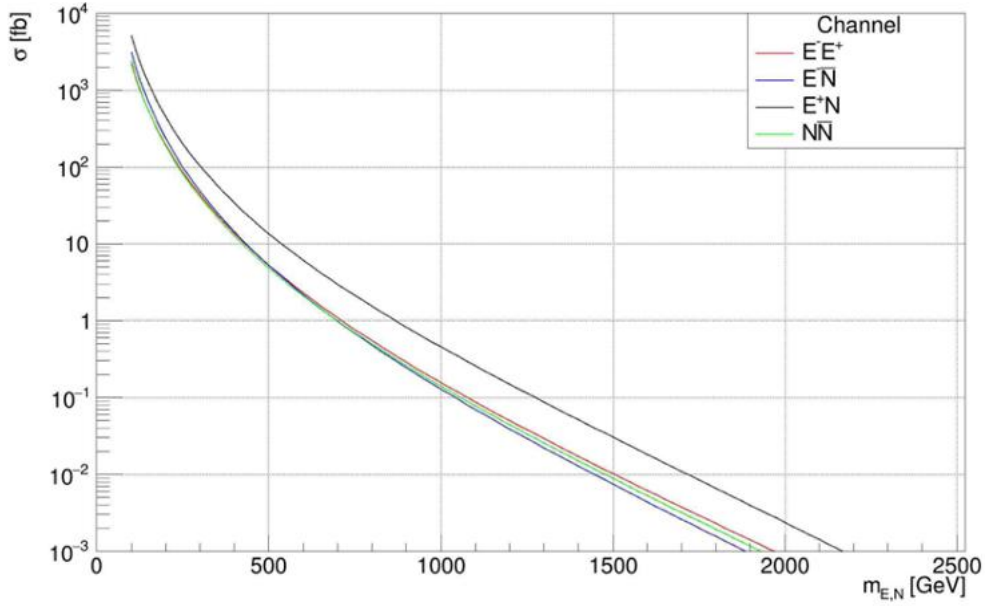

**Figure 5:** Cross-sections for pair production of iso-doublet VLLs at the LHC with $\sqrt{s}$ = 13 TeV.

Let us emphasize that:

i) The assumption of degeneracy is valid if only one vector-like lepton doublet exists. In much more natural case each SM family should have its own VLL doublet, as in $E_6$ GUT. Therefore, all three production channels should be analyzed separately,

ii) For the iso-doublet case, decay channels induced by the existence of $\nu_R$ should be taken into account,

iii) The unsuppressed decay channel E → NW (N → EW), which may be dominant if $M_E > M_N$ ($M_N > M_E$), should not be ignored.

Therefore, ATLAS and CMS *analyses should be redone keeping in mind the missing decay channels for each production channel as well.*

*Conclusions.—* It is seen that a number of possible decay channels of charged and neutral, iso-singlet and iso-doublet vector-like leptons have been omitted in experimental analyses carried out so far. Two reasons for disregarding these channels are the degenerate masses of charged and neutral VLLs in the RM and the assumption of the absence of right-handed neutrinos. While the former may be true if only one VLL pair is present, neutrino oscillations disprove the latter.

Recent ATLAS and CMS analyses are concentrated on the third SM family lepton partners. This is a consequence of the assumption that only one iso-doublet exists and the incorrect application of the CKM structure to vector-like leptons (for details, see [15]). *Therefore, the first and second SM family leptons' VLL partners should be analyzed as well.* Let us repeat that the Restricted Model, which was used in ATLAS and CMS analyses at the 13 TeV center of mass energy, does not include a number of possible decay channels. Moreover, these channels may be dominant depending on mass differences and mixing angle values.

Concerning L3 results, they should be reanalyzed, taking into account additional decay modes. Especially, analysis of pair production of neutral VLLs with subsequent N → Zν decays may lead to surprising results (this statement may also be correct for LHC). It should be underlined that the exclusion limits obtained by L3, ATLAS and CMS Collaborations are valid for their specific analyses and cannot be considered as general exclusion limits.

Finally, there are strong phenomenological arguments for the existence of vector-like leptons and corresponding experimental studies should be strengthened. Let us emphasize that (as mentioned in [15]) "Discovery of VLLs and VLQs will shed light on the mass and mixing patterns of fundamental (at today's level) fermions." As pointed out in the Introduction, several years ago Steve Weinberg stated that the observed pattern of quark and lepton masses is one of the most important mysteries to be solved [3].